# Synthesis of metalloborophene nanoribbons on Cu(110)

*Xiao-Ji Weng, Yi Zhu, Ying Xu, Jie Bai, Zhuhua Zhang, Bo Xu\*, Xiang-Feng Zhou\*, Yongjun Tian*

X.-J. Weng, Y. Zhu, J. Bai, B. Xu, X.-F. Zhou, Y. Tian
Center for High Pressure Science, State Key Laboratory of Metastable Materials Science and Technology, School of Science, Yanshan University, Qinhuangdao 066004, China
E-mail: xfzhou@ysu.edu.cn; bxu@ysu.edu.cn

Y. Xu, Z. Zhang
Key Laboratory for Intelligent Nano Materials and Devices of Ministry of Education, State Key Laboratory of Mechanics and Control of Mechanical Structures, Institute for Frontier Science, Nanjing University of Aeronautics and Astronautics, Nanjing 210016, China

Metalloborophene, characterized by the presence of metal-centered boron wheels denoted as M©$B_n$, has garnered considerable attention in recent years due to its versatile properties and potential applications in fields such as electronics, spintronics, and catalysis. However, the experimental verification of metalloborophene has been challenging, mainly due to the unconventional two-dimensional (2D) boron networks. In this study, we employ scanning tunneling microscopy, X-ray photoelectron spectroscopy, low energy electron diffraction, and first-principles calculations to unveil Cu©$B_8$ metalloborophene nanoribbons formed via spontaneous alloying after the deposition of boron on a heated Cu(110) substrate under ultrahigh vacuum condition. The thermodynamically preferred precursor, the anchoring of boron network to metal atoms, and anisotropic lattice mismatch are identified as pivotal factors in the formation of these metalloborophene nanoribbons. This discovery expands the repertoire of 2D materials and offers a potential pathway for the synthesis of other metalloborophenes.

## 1. Introduction

Metalloborophenes has emerged as a novel class of two-dimensional (2D) materials that comprise a boron-rich framework and embedded metal atoms,[1,2] promising planar hyper-coordinate chemical bonding possibilities.[3] The incorporation of variant metal atoms in different configurations holds the potential to tailor properties such as catalytic, magnetic, and optical behaviors. For instance, $FeB_6$ monolayers, featuring either Fe©$B_8$ or Fe©$B_6$ units, exhibit metallic or semiconductor characteristics with remarkable visible-light absorption capabilities.[4] Additionally, a ferromagnetic $CrB_4$ nanosheet with the Cr©$B_8$ motif has been predicted to possess a Curie temperature exceeding room temperature.[5] Furthermore, several metalloborophenes containing honeycomb boron sheets doped with metals such as Fe, Zr, Ti, Hf, V, Nb, and Ta have been anticipated to display Dirac band dispersion or in-plane negative Poisson's ratio.[6–9]

While the constituents of metalloborophene, namely metal-doped boron clusters, were proposed earlier and subsequently verified through experiments,[10–15] the lack of experimental evidence for the existence of metalloborophene can be attributed, in part, to the unconventional 2D boron networks and the absence of a parent bulk phase. As such, achieving metalloborophene through methods such as mechanical cleavage or liquid phase exfoliation remains a significant challenge. Alternatively, epitaxial growth presents a viable approach for synthesizing metalloborophene on a suitable substrate. Recently, the successfully synthesis of a 2D copper boride consisting of atomic zigzag Cu and B chains on Cu(111) surface has demonstrated the feasibility of attaining unconventional 2D boron networks.[16–19] In this work, we managed to synthesize metalloborophene on Cu(110) substrate with molecular beam epitaxy. After boron deposition, parallel nanoribbons along Cu[-110] were developed on the substrate. The nanoribbons were characterized using scanning tunneling microscopy (STM), X-ray photoelectron spectroscopy (XPS), low energy electron diffraction (LEED), and first-principles calculations. This combinatory characterization resolved the Cu©$B_8$ structural motif, indicating the successful formation of metalloborophene. Preliminary investigations into the eletronic structure, catalytic properties, and magnetism of the metalloborophenes further revealed the potential for diverse applications.

## 2. Results and Discussion

### 2.1. Structure of metalloborophene

The Cu©$B_8$ nanoribbon was synthesized by depositing pure boron on a clean Cu(110) surface under ultrahigh-vacuum condition. The substrate was heated to 430°C and subsequently exposed to boron beam before cooling to room temperature. Auger electron spectroscopy (AES) confirmed the successful growth of a boron-based epitaxial structure, as indicated by the appearance of the B KLL peak after deposition (**Figure 1**a). X-ray photoelectron spectroscopy (XPS) revealed a minor shift of 0.1 eV for the Cu $2p^{3/2}$ core level towards the higher binding energy after boron deposition (Figure 1b). It implies the existence of copper-boron interaction, leading to the presence of a positively charged surface copper component. The B 1s core level displayed a single peak at 187.7 eV (Figure 1c), distinct from the multiple peaks typically associated with borophene.[20,21] This observation suggests that surface boride formation occurs via spontaneous alloying on the heated substrate.

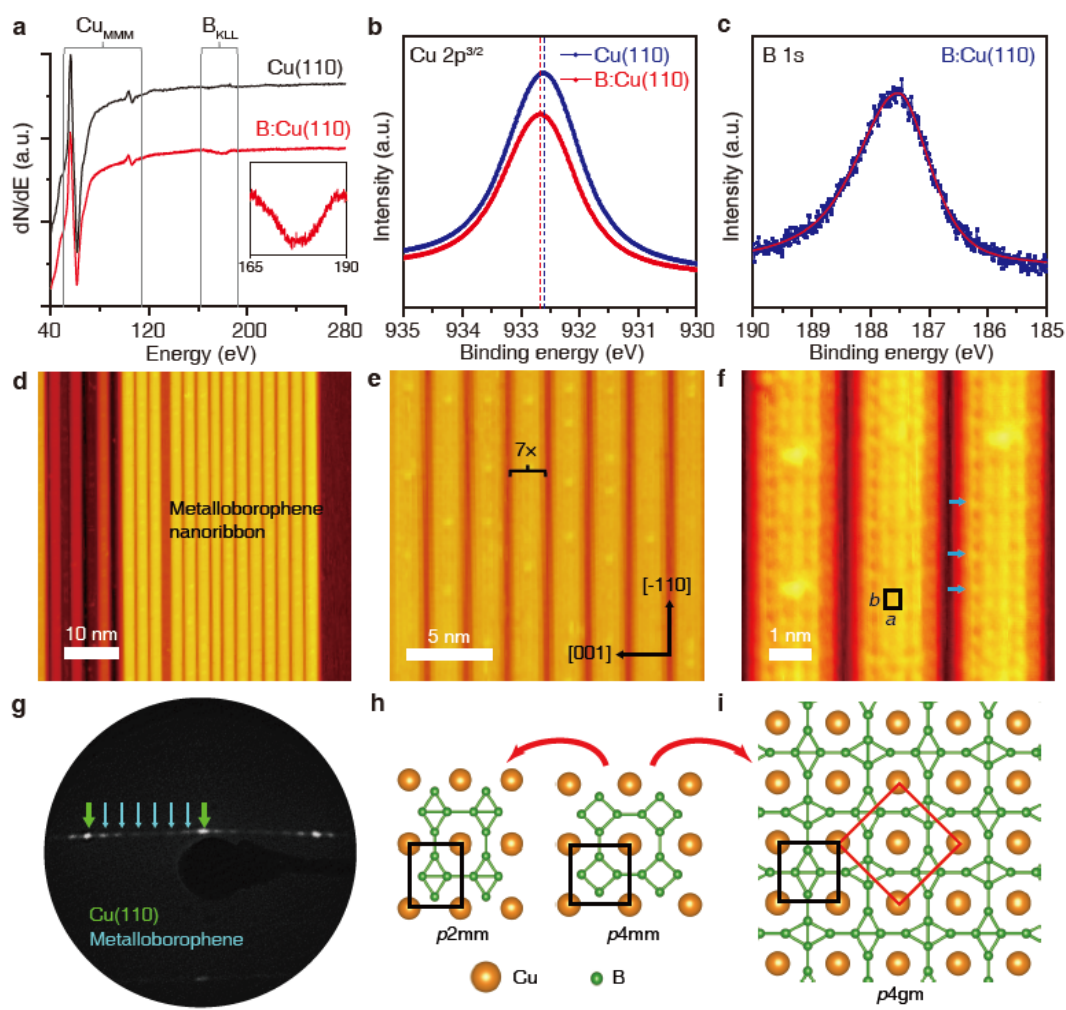

**Figure 1.** Structure characterization of the metalloborophene nanoribbon. a) AES data from Cu(110) before and after boron deposition. b) and c) XPS data of Cu $2p^{3/2}$ and B 1s core levels. d) Representative STM image of the boron-deposited Cu(110) ($I_t$ = 250 pA, $V_t$ = 0.782 V). e) STM image of homogeneous metalloborophene nanoribbons ($I_t$ = 330 pA, $V_t$ = 0.911 V). f) High-resolution STM image of the nanoribbons as in (e) ($I_t$ = 300 pA, $V_t$ = 0.911 V). g) LEED patterns of Cu(110) dominated with metalloborophene nanoribbons, taken with a beam energy of 41 eV. h) CuB$_4$ sheet with planar symmetry of *p*4mm and *p*2mm, respectively. i) Atomic configuration of the proposed *p*4gm-CuB$_4$ monolayer.

Scanning tunneling microscopy revealed a characteristic surface morphology comprising multiple nanoribbons (Figure 1d), all oriented parallel to Cu[-110]. Notably, the nanoribbons, as bright stripes with different widths and heights, did not manifest along other directions on Cu(111) surface, underscoring their highly anisotropic epitaxial growth. Among the observed nanoribbons, our study focused on those with a relatively large width and high coverage, approximating a quasi-two-dimensional form. These nanoribbons exhibited a width of *ca.* 2.53 nm, roughly seven times the lattice constant of Cu(110) along the [001] direction (*i.e.*, 0.361 nm). Depending on the amount of deposited boron, a dominated 7× surface phase was successfully achieved, as evidenced by the sevenfold signal in the low-energy electron diffraction (LEED, Figure 1g) pattern. Figure 1e shows a surface region containing a single domain of homogeneous nanoribbons with a 7× periodicity. Without significant impurities and defects, dense grid-like pattern was discernible within the nanoribbon structure. High-resolution STM image unveiled a series of sequenced protrusions contributing to these patterns (Figure 1f). These protrusions are arranged along the [001] and [-110] directions, featuring lattice parameters of $a = 0.38$ nm and $b = 0.41$ nm. Importantly, this lattice structure does not conform to a typical borophene structure that is characterized by the regular triangle boron sheet with B-B bond length of *ca.* 0.17 nm. Consequently, the nanoribbons are most likely a result of the reaction between the Cu(110) substrate and the deposited boron.

Considering the parameters of the unit cell and the arrangement of protrusions, a stoichiometric estimate suggests the presence of one $CuB_4$ formula per unit cell. Alternative stoichiometries, such as $Cu_2B_x$, $CuB_3$, and $CuB_5$, were explored in trials but led to substantial mismatches. In light of the near-square lattice configuration, a preliminary structure model comprising planar $B_4$ clusters and Cu atoms, denoted as *p*4mm-$CuB_4$, was proposed (Figure 1h), where every Cu atom is surrounded by eight B atoms, featuring the Cu©$B_8$ motif. However, first-principles calculations indicated that the $B_4$ cluster preferred to exist in a diamond shape consisting of two $B_3$ triangles, resulting in a transformation from *p*4mm-$CuB_4$ to *p*2mm-$CuB_4$ with a rectangular lattice of $a = 0.360$ nm and $b = 0.446$ nm. Notably, lattice parameters *a* and *b* of *p*2mm-$CuB_4$ deviates from the corresponding experimental values by -5.3% and 8.8%, respectively. To address this discrepancy, a more reasonable model was proposed, wherein the nearby $B_4$ diamonds were oriented perpendicular to each other. This adjustment yields a distance of 0.396 nm between two nearest Cu atoms, which is 3.9% longer in the [001] direction and 3.6% shorter in the [-110] direction compared to the measured values (Figure 1i). Given the reduced lattice mismatch in comparison to the experimental results, the atomic structure of *p*4gm-$CuB_4$ (Cu©$B_8$) was ultimately adopted for further investigation. Intriguingly, the structural feature in this model aligns with previously predicted structures of *P*4/*MBM* CrB$_4$ or MnB$_4$.[5]

To validate the proposed structure model and gain a comprehensive understanding of the anisotropic growth behavior of the metalloborophene, it is essential to analyze the periodicities along the two directions of Cu[-110] and [001]. Although the primitive cell (see the red box in Figure 1i) of *p*4gm-CuB$_4$ encompasses two CuB$_4$ formulas, for the sake of simplicity, we consider the unit cell containing a single Cu atom (see the black box in Figure 1i) as a 1×1 unit in our subsequent discussion.

In the [001] direction, the measured period of the surface structure, amounting to 0.38 nm, closely approximates that of Cu(110) substrate, which is 0.361 nm. Conversely, the measured period of 0.41 nm along the [-110] direction exhibits a significant mismatch with the Cu(110) lattice constant, which is 0.256 nm. Consequently, we propose three configurations to assess the lattice matching (**Figure 2**a), *i.e.*, 1×2 CuB$_4$ on 1×3 Cu(110), 1×3 CuB$_4$ on 1×5 Cu(110), and 1×5 CuB$_4$ on 1×8 Cu(110). The mismatch in the [-110] direction for these three configurations is -6.3%, 4.1%, and -0.1%, respectively. Note that the last configuration is a combination of the first two, effectively eliminating any residual tension or compression along the [-110] direction (0.256×8 ≈ 0.41×5). In contrast, relieving the 5.3% mismatch in the [001] direction would require a significantly longer length of 6.84 nm (0.361×19 ≈ 0.38×18). This underscores the key role of a better matching along [-110] in the formation of the nanoribbons.

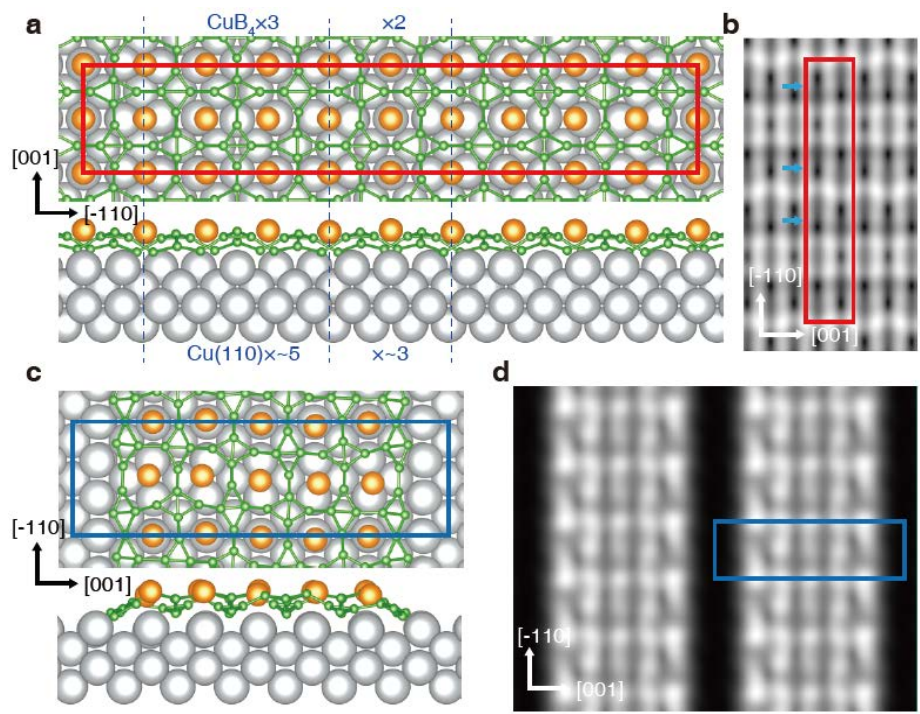

**Figure 2.** Structure of the Cu©B$_8$ metalloborophene. a) Top and side view of 2×10 CuB$_4$ on 2×16 Cu(110). b) Simulated STM image of the structure in (a). c) Top and side view of one

Cu©B$_8$ nanoribbon with terminated boron edges. The red box marks a 7×3 unit cell of Cu(110) surface. d) Simulated STM image of the model in (c). Orange, green, and gray spheres represent the surface copper atoms, boron atoms, and substrate copper atoms, respectively.

Figure 2a illustrates the structure model, featuring 2×10 CuB$_4$ on 2×16 Cu(110). Following a thorough relaxation process through the density functional theory (DFT) calculation, the boron network exhibited a spontaneous out-of-plane buckling, and the Cu atoms appeared to be elevated (Figure S1). Importantly, this process did not induce substantial fluctuation or collapses across the surface, consistent with the flat extension of the nanoribbon along the [-110] direction. To gain insights into the surface charge distribution, we simulated the STM image by integrating states near the Fermi level for the modelling structure. A comparison between Figure 2a and 2b confirmed that the protrusions in the STM pattern primarily originated from the topmost Cu atoms in the metalloborophene. Furthermore, the charge density appeared to be modulated by the substrate in double or triple period of CuB$_4$, in agreement with the brightness modulation as indicated by the blue arrows in Figure 1f. This modulation is likely associated with the matching configuration of 1×2 CuB$_4$ on 1×3 Cu(110) and 1×3 CuB$_4$ on 1×5 Cu(110).

Based on the observation of five-column protrusions within each nanoribbon (Figure 1e and 1f), we constructed a structure consisting of five periods of CuB$_4$ in the [001] direction of Cu(110) surface (Figure 2c). After relaxation, the surface structure exhibited no substantial distortions. A notable change in the nanoribbon structure was the boron edges along the [-110] direction were firmly anchored to the Cu(110) substrate. The simulated STM image (Figure 2d) reveals the presence of five-column protrusions, which generally aligns with the positions of surface Cu atoms. The average distance between nearby columns is approximately 0.38 nm, in good agreement with the experimental findings.

**2.2. Precursor of metalloborophene**

Understanding the formation mechanism of such an unconventional boron network is of paramount importance, as it can pave the way for the realization of other metalloborophenes in the future. Through the simulation of the epitaxial structures and a deeper exploration of the substructure of CuB$_4$, we propose a structure of boron nanochain on Cu(110) as the plausible precursor. This prediction is substantiated by the formation energy ($E_f$) of -6.68 eV/atom, notably lower than that of α-sheet borophene,[22] *i.e.*, -6.28 eV/atom. As illustrated in **Figure 3**a, this one-dimensional nanostructure comprises B$_4$ diamonds and B-B dimers, recognized as the primary constituent of CuB$_4$ (as indicated by the blue and purple substructures in Figure 3b). The neighboring nanochains are then interconnected by individual boron atoms, ultimately resulting in the formation of the unconventional boron network (Figure 3c). What is particularly striking is the presence of octagonal boron rings within these assemblies, which possess the capability to accommodate Cu atoms from the substrate, consequently leading to the formation of a Cu-centered boron wheel structure. This observation suggests that the metalloborophene nanoribbons bearing the Cu©B$_8$ motif may emerge through the expansion of boron nanochains in conjunction with the interaction with surface Cu atoms. However, due to the absence of strict *in-situ* observation for the growth process, it cannot be excluded that there is other intermediate product contributing to the generation of metalloborophene. For example, some borozenes are formed firstly and then react with the copper atoms, finally result in the formation of one-dimensional nanostructures.

As demonstrated in Figure 3e, large areas of 3× nanochain structure can be synthesized over Cu(110) by lowering the boron deposition dosage. The triple periodicity along the [001] direction is further confirmed by the LEED pattern (Figure 3d), which is similar to a recent report.[23] Furthermore, Figure 3e indicates that even on a surface dominated by the 3× phase, the 7× phase characterized by the Cu©B$_8$ motifs can appear occasionally. This observation strongly evidences that the two phases can coexist on the Cu(110) substrate. While the precise structure of the 3× phase remains to be fully elucidated, it is highly likely to be a derivative of boron nanochains (Figure 3a and 3f).

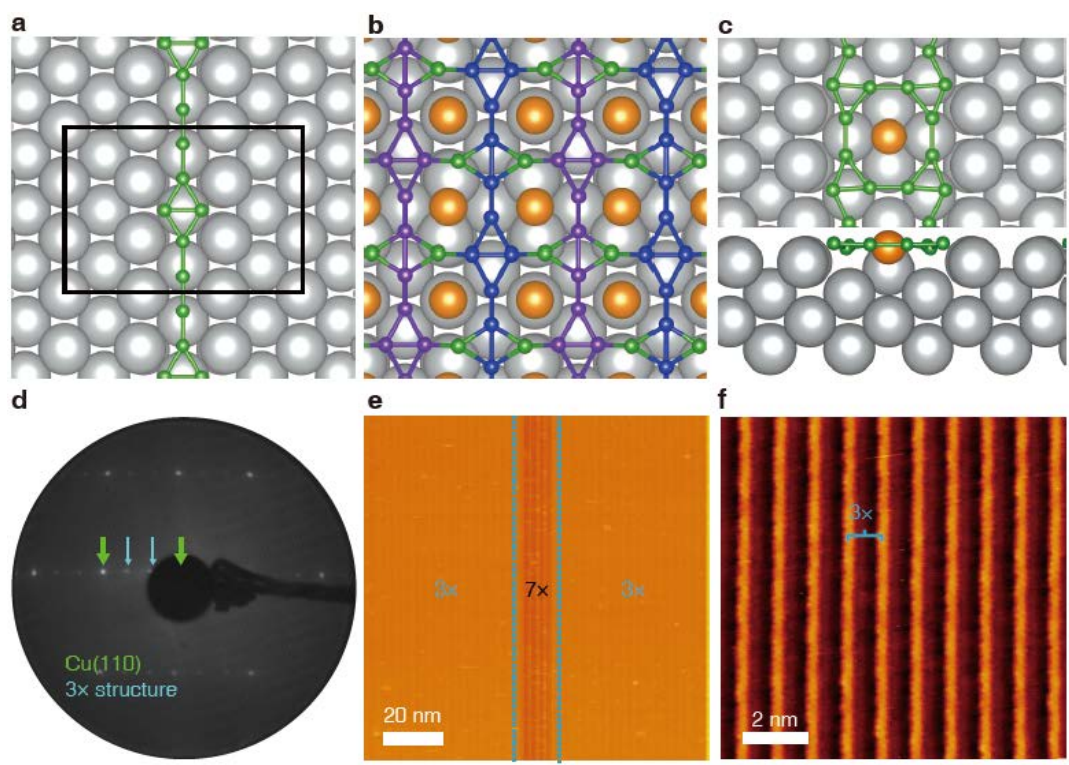

**Figure 3.** Possible precursor for the metalloborophene. a) Structure of a boron nanochain. b) The assembled metalloborophene on Cu(110). c) The surface Cu atoms (colored in orange)

were pulled outwards by two connected boron chains. d) LEED pattern for the sample prepared with lower boron dosage, taken at a beam energy of 120 eV. e) STM image for the sample prepared with lower boron dosage. f) High-resolution STM image of the 3× phase.

## 2.3. Properties of metalloborophene

We conducted the electronic properties calculations of metalloborophene both with and without the substrate. It is indicated that CuB$_4$ exhibits metallic behavior on Cu(110), as well as in the freestanding form (**Figure 4**a and Figure S2). This behavior is attributed to the strong hybridization of Cu-*d* and B-*p* orbitals, resulting in a substantial mixing of states in the vicinity of the Fermi level, as shown in Figure 4b. Note that interaction between copper and boron atoms can be covalent or ionic in different CuB$_n^-$ clusters.[24–26] Here the nearest Cu-B distance in *p*4gm-CuB$_4$ is 2.04 Å, longer than a typical value of Cu-B single bond (1.97 Å),[24] implying the covalent interaction between copper and boron is not strong. To delve deeper into the electronic structure, we performed Bader charge analysis, an approach that provides insight into the charge distribution around each atom by utilizing zero flux sheets of electron density.[27] Our calculation reveals a Bader charge of +0.36$e$ for the Cu atoms within Cu©B$_8$, indicating an ionic interaction between the central Cu atoms and the octagonal boron rings. On the other hand, the average electron accumulation on the surface boron amounted to -0.16$e$ per atom, contributed from both the central Cu atoms and the Cu(110) substrate. This charge transfer between the substrate and the metalloborophene, as shown in Figure 4c, is reminiscent of the behavior observed in borophene on metal surface.[28] Furthermore, the exfoliation energy of CuB$_4$ was estimated to be about 1.25 J/m$^2$, which falls below the exfoliation energy criterion of 2.08 J/m$^2$, indicative of the potential for exfoliation.[29]

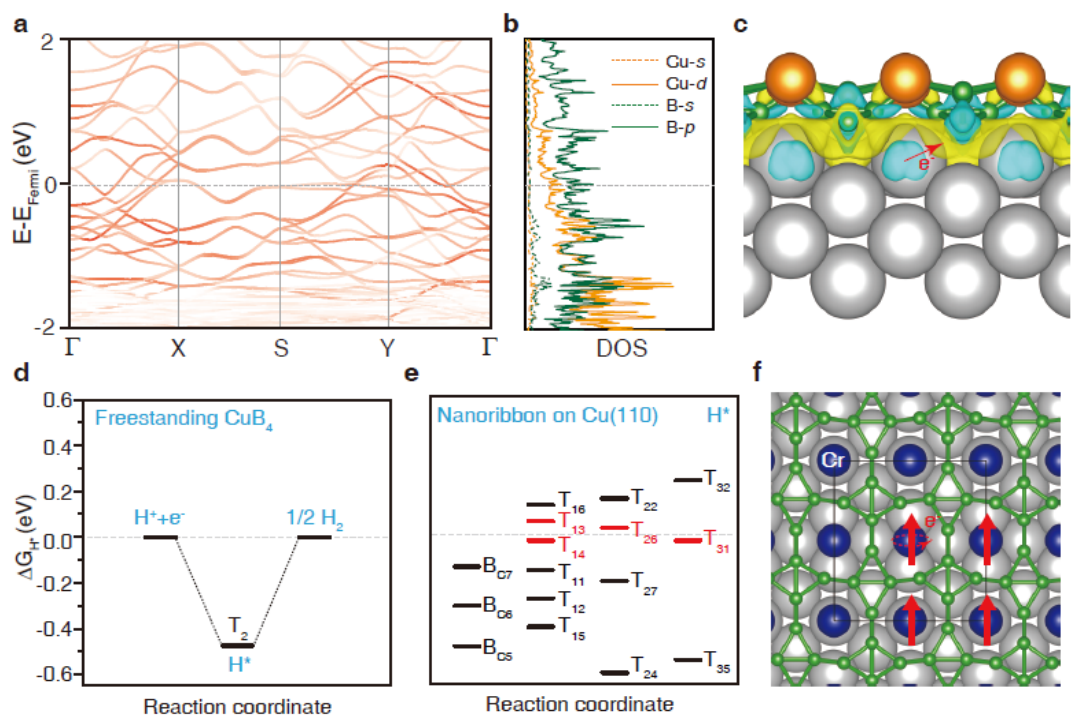

**Figure 4.** Electronic properties of the metalloborophene. a) Band structure of Cu©B$_8$ metalloborophene on Cu(110). The shade of red color represents the projected weight of the electron states from the metalloborophene. b) Density of states (DOS) of Cu©B$_8$ metalloborophene on Cu(110). The curves show the orbital-projected DOS from copper and boron. c) Charge density difference between the whole structure and the separated metalloborophene and substrate. d) and e) Computed Gibbs free energies for hydrogen adsorption on freestanding metalloborophene and metalloborophene nanoribbon on Cu(110), respectively. f) The magnetic structure of Cr©B$_8$ on Cu(110).

To explore the material's electrocatalytic potential, we simulated the hydrogen evolution reaction by calculating various H-adsorbed configurations on CuB$_4$. In the case of freestanding CuB$_4$, hydrogen only adsorbs on one of the B atoms, with the H-adsorption site denoted as T$_2$, bearing a Gibbs free energy of -0.474 eV (Figures 4d and S3a). In contrast, the metalloborophene nanoribbon on Cu(110) exhibited multiple non-equivalent H-adsorption sites exist (Figures 4e and S3b). The Gibbs free energy associated with hydrogen adsorption at these sites ranges from -0.601 to 0.235 eV, while for sites T$_{31}$, T$_{14}$, T$_{26}$ and T$_{13}$, it was calculated to be -0.025, -0.026, 0.029, and 0.058 eV, respectively. Notably, compared with 0.104 eV for the bare Cu(110) surface and *ca*. 0.1 eV for the noble metal Pt,[30,31] the metalloborophene nanoribbon on Cu(110) displayed superior catalytic performance in the hydrogen evolution reaction. Furthermore, to explore potential avenues for tuning the material's electronic properties, we replaced copper in Cu©B$_8$ with various transition metal elements. Our results indicate that certain structures can retain the M©B$_8$ configuration, such as Cr@B$_8$ (Figure 4f).

First-principles calculations predicted the existence of ferromagnetic CrB$_4$ and MnB$_4$ on Cu(110), consistent with their freestanding counterparts.[5]

## 3. Conclusion

In summary, metalloborophene nanoribbons were successfully synthesized through a spontaneous reaction between the deposited boron and the Cu(110) substrate. Within this novel structure, individual copper atoms are incorporated into the boron wheels, featuring the distinctive Cu©B$_8$ motif. This motif serves as a prototype for the envisaged MB$_4$ (M = Mn, Cr, and so forth) sheets. The formation of the metalloborophene nanoribbons depends on three key factors: the thermodynamically preferred precursor of boron nanochains, the accommodation of metal atoms in the boron network, and the anisotropic lattice mismatch. DFT calculations further suggest potential applications in catalysis and spin electronics for these M©B$_8$ structures. This discovery paves the way for future research on a variety of metalloborophenes, each with distinct mosaic patterns and offering opportunities to finely tailor the chemical and physical properties.

## 4. Methods

*Growth and characterizations*: A single crystal Cu(110) substrate was placed into a ultrahigh-vacuum (UHV) system (base pressure: $2\times10^{-10}$ mbar) with the interconnected growth and characterization chambers. A clean and smooth Cu(110) was obtained *via* repeated cycles of Ar$^+$ sputtering and annealing. The pure boron powder in the pyrolytic graphite crucible was evaporated by electron-beam heating, while the substrate was kept at about 430 °C (measured by a thermocouple). After growth and cooling, the sample was transferred to the characterization chamber under UHV condition and imaged with STM at room temperature (Aarhus VT-STM). LEED patterns were collected in an electron energy range of 20–300 eV (SPECS ErLEED). Auger electron spectroscopy (AES) was taken with an incident electron energy of 1000 eV. X-ray photoelectron spectroscopy (XPS) was performed with Al Kα radiation.

*Computational methods*: First-principles calculations were performed by employing the density functional theory (DFT) as implemented in the VASP code.[32] Projector-augmented wave (PAW) methods and generalized gradient approximation (GGA) with the functional of Perdew, Burke, and Ernzehof (PBE) were adopted.[33,34] Plane-wave cutoff energy of 450 eV, Γ-centered $k$-point grid with a resolution of $2\pi \times 0.04$ Å were chosen to ensure that the energy convergences and the force are smaller than $10^{-6}$ eV and 0.01 eV/Å, respectively. The STM images were simulated by integrating the electronic state near the Fermi level based on the Tersoff-Hamann theory and visualized by p4VASP program.[35] The tunneling bias voltage (V$_t$) was set to 1.0 eV for the simulation, being close to the experimental condition. The formation energy of the boron structure was calculated by

$$E_f = (E_{tot} - E_{sub})/N_B \tag{1}$$

Where the $E_{tot}$ and $E_{sub}$ are the energies of the B/Cu system and the Cu(110) substrate, respectively, $N_B$ is the number of B atoms. The vaspkit code and VESTA software were utilized for the analysis and illustration of the electronic properties.[36,37] The strongly-correlated correction is calculated with GGA+U method[38] to treat the $3d$ electrons of transition metal elements. The adsorption energy of hydrogen is defined as

$$\Delta E_H = E_{surf+nH} - E_{surf} - n_H E_{H_2}/2 \tag{2}$$

Where $E_{surf+nH}$, $E_{surf}$ and $E_{H_2}$ are the energies of the H-adsorbed B/Cu system, the pristine B/Cu system (including the freestanding metalloborophene and its nanoribbon on Cu(110)), and hydrogen molecule $H_2$, respectively. $N_H$ is the number of the adsorbed H atoms. The Gibbs free energy of the adsorbed state was estimated as

$$\Delta G = (\Delta E_H + \Delta E_{ZPE} - T\Delta S_H + \Delta E_{solv})/n_H \tag{3}$$

where $\Delta E_{ZPE}$ is the difference of zero-point energy between the adsorbed and the gas phase, while $\Delta S_H$ is the change in entropy at room temperature ($T = 300$ K). The parameter of $E_{solv}$ is the solvation energy that evaluated using the VASPsol implementation of the implicit solvation model.[39]

**Supporting Information**

Supporting Information is available from the Wiley Online Library or from the author.

**Acknowledgements**

This work was supported by the National Natural Science Foundation of China (Grants 52025026, 22073048, 12272180), the Fundamental Research Funds for the Central Universities (NS2022003), and the Natural Science Foundation of Hebei Province of China (E2022203109).

Metalloborophene nanoribbons characterized by the metal-centered boron wheels (M©B$_8$) is synthesized by depositing boron onto copper surface under ultrahigh vacuum condition. Scanning tunneling microscopy, X-ray photoelectron spectroscopy and first-principles calculations revealed several crucial factors contributing to the formation of this unexpected 2D material that hold promise for tailoring various properties such as catalysis and magnetism.

X.-J. Weng, Y. Zhu, Y. Xu, J. Bai, Z. Zhang, B. Xu*, X.-F. Zhou*, Y. Tian

**Synthesis of metalloborophene nanoribbons on Cu(110)**

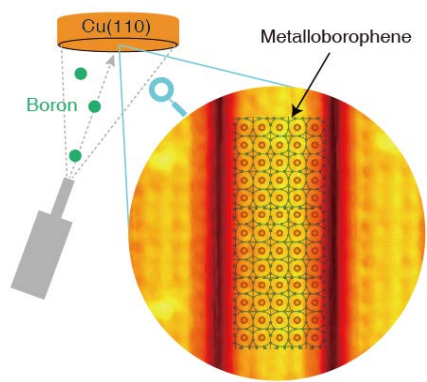

# Supporting Information

**Synthesis of metalloborophene nanoribbons on Cu(110)**

*Xiao-Ji Weng, Yi Zhu, Ying Xu, Jie Bai, Zhuhua Zhang, Bo Xu\*, Xiang-Feng Zhou\*, Yongjun Tian*

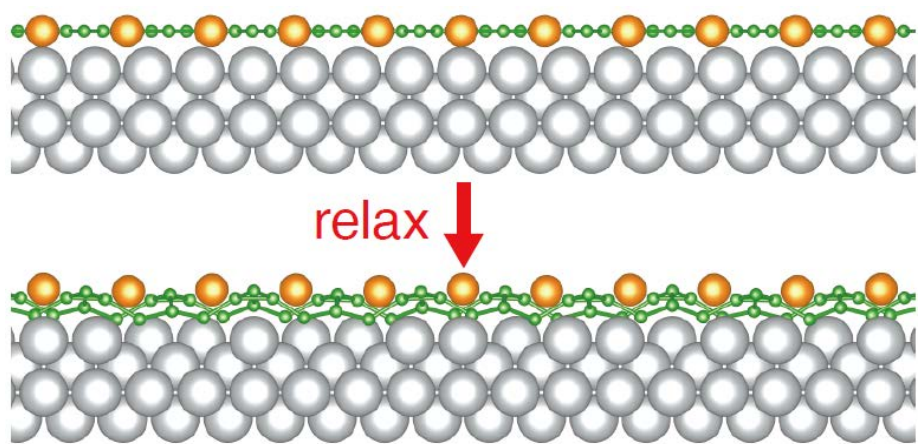

**Figure S1** Side view of the atomic configuration of 2×10 $CuB_4$ on 2×16 Cu(110) before and after DFT relaxation.

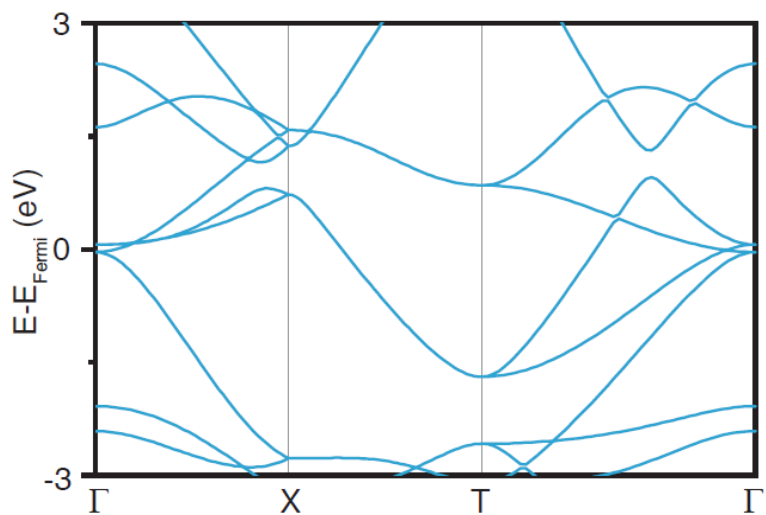

**Figure S2** Calculated band structure of the freestanding *p*4gm-$CuB_4$.

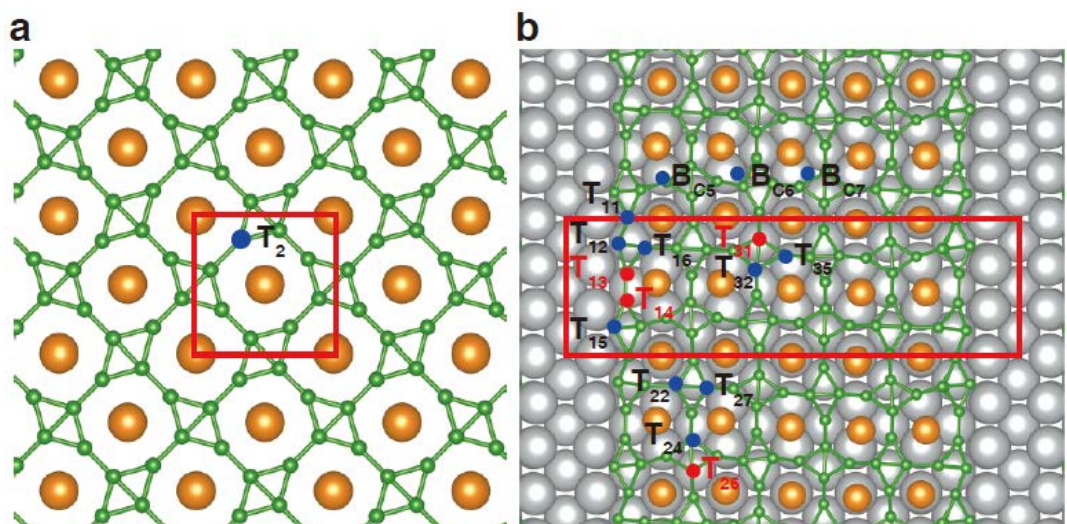

**Figure S3** The hydrogen adsorption sites on Cu©B$_8$ metalloborophene. **a**, Freestanding *p*4gm-CuB$_4$. **b**, Metalloborophene nanoribbon on Cu(110).